\documentstyle[aps,12pt,manuscript]{revtex}
\begin{document}
\preprint{INJE-TP-98-4}

\title{Greybody factor for the BTZ black hole and a 5D black hole}

\author{ H.W. Lee and Y. S. Myung}
\address{Department of Physics, Inje University, Kimhae 621-749, Korea} 

\maketitle

\begin{abstract}
We study the 5D black holes in the type IIB superstring theory 
compactified on $S^1 \times T^4$. Far from horizon, we have 
flat space-time. Near horizon, we have 
$AdS_3({\rm BTZ~ black~ hole}) \times S^3 \times T^4$. 
We calculate the greybody factor  
of a minimally coupled scalar by replacing the original geometry(
$M_5 \times S^1 \times T^4$) by $AdS_3 \times S^3 \times T^4$. 
In the low-energy scattering, it turns out that 
 the result agrees with the greybody factor of the 5D black hole 
(or D1 + D5 branes)in the 
dilute gas approximation.
This confirms that the $AdS$-theory($AdS_3 \times S^3 \times T^4$) 
contains the essential information about 
the bulk 5D black holes.
\end{abstract}

\newpage
\section{Introduction}
Recently anti-de Sitter spacetime($AdS$) has attracted much interest. It
appears that the conjecture relating
the string(supergravity) theory on $AdS$ to conformal field theory(CFT) 
on its boundary
may resolve many problems in black hole physics\cite{Mal97,Gub98,Wit98,Hor98}.
The 5D black hole correspond to U-dual to
BTZ$\times S^3$\cite{Hyu97,Sfe97}. 
Sfetsos and Skenderis calculated the entropies
of non-extremal 5D black holes by applying Carlip's
approach to the BTZ black hole. The BTZ black hole(locally $AdS_3$)
has no curvature singularity\cite{Ban92}
and is considered as a prototype for the general CFT/$AdS$ correspondence.
This is actually an exact solution of string theory
\cite{Hor93}
and there is an exact CFT with it on the
boundary. Carlip has shown that the physical boundary
degrees of freedom account for the Bekenstein-Hawking
entropy of the BTZ black hole correctly\cite{Car95}.
Another approach to calculate the entropy of the BTZ black hole 
was discussed in \cite{Str97}.

Apart from counting the microstates of black holes, the dynamical 
behavior is also an important issue\cite{Dha96}. This is 
so because the greybody factor for the black hole arises 
as a consequence of scattering of a test field off the 
gravitational potential barrier surrounding the horizon. 
That is, this is an effect of spacetime curvature. Together 
with the Bekenstein-Hawking entropy, this seems to be a 
strong hint of a deep and mysterious connection between 
curvature and statistical nechanics. It was shown that the 
s-wave greybody factor for the BTZ black hole has the same 
form as the 5D black hole in the dilute gas approximation\cite{Bir97}. 

Here we wish to calculate the greybody factors, using a minimally 
coupled scalar. The first one is based on 
the bulk supergravity calculation\cite{Dow97,Mat97,Mal97-1} 
in the effective 5D black hole background. 
This may be considered as a review of past works. But some 
discrepancy appears because of the extra dimensions($T^4$). 
Also the D-brane results (effective CFT calculations) are
briefly reviewed for comparison. 
The second is performed by replacing the original geometry 
($M_5 \times S^1 \times T^4$) by $AdS_3 \times S^3 \times T^4$.
But we do not require any boundary condition(Dirichlet or Neumann) 
at the spatial infinity, which is necessary for studying the 
whole $AdS$. 

\section{Geometry}
The 5D black hole can be obtained as a compactified solution 
on $S^1 \times T^4$ to 
the low-energy effective action of the 
type IIB theory\cite{Hor96}. 
This is also represented by the D-brane picture. 
The near-extremal 5D black hole is described by 
the bound states of $Q_1$ 1D-branes and 
$Q_5$ 5D-branes with some momentum flowing along the 
1D-brane. Non extremality is achieved by introducing both left and 
right momenta on the 1D-brane. The 
supergravity solution in the string frame is given by
\begin{eqnarray}
ds_{10}^2 &=& (f_1 f_5)^{-1/2} \left [ -dt^2 + dx_5^2 +
{r_0^2 \over r^2} ( \cosh \sigma dt - \sinh \sigma dx_5)^2 +
f_1 dx_i^2 \right ]
\nonumber \\
&&
~~~~~~~~~~~~~~
+(f_1f_5)^{1/2} \left [ h^{-1} dr^2 + r^2 d\Omega_3^2 \right ],
\label{string-metric}
\end{eqnarray}
where $f_1 = 1 + r_1^2 / r^2, f_5 = 1 + r_5^2 / r^2, 
h= 1 - r_0^2 / r^2$. The various length scales are given by 
$r_1^2 = g Q_1 \alpha'^3/V, r_5^2=g Q_5 \alpha', r_0^2 \sinh 2\sigma =
2 g^2 n \alpha'^4 /R^2 V$, and $r_n^2 = r_0^2 \sinh^2 \sigma$. 
Here $V(\sim \alpha'^2)$ is 
the volume of $T^4$. The integers 
$Q_1, Q_5, n$ are charges 
for 1D, 5D-branes and the total momentum. 
$g$ is the string coupling and $R$ is the radius of $S^1$. 
The brane configuration lies on a $S^1 \times T^4$ with 1D-brane 
along $S^1$. 

In fact the semiclassical black hole picture 
and the perturbative D-brane picture are considered to describe the same object 
in two different regimes. The D-brane picture is expected to 
describe the semiclassical black hole when $gQ$ 
is large, whereas
the D-brane calculations are performed at weak coupling(
$gQ \ll1$). For extremal BPS states there are well-known 
non-renormalization theorems which state that the degeneracy of states 
do not change as $gQ$ increases. However, for non-BPS states including 
the dilute gas limit, there are no such theorems. One wishes to 
resolve this puzzle with a new idea.
For this purpose, one may  
introduce the decoupling limit which should be taken to suppress 
closed string loop corrections($g \to 0$) and higher-order terms 
in the $\alpha'$-expansion ($\alpha' \to 0$)\cite{Mal97}. This 
is given by
\begin{equation}
g \to 0, 
\alpha' \to 0  ; g Q_1, g Q_5 \gg 1 ~~{\rm with}~~ 
r_1, r_5, r_n ~{\rm fixed}.
\label{decoupling-limit}
\end{equation}
Actually (\ref{decoupling-limit}) 
corresponds to the dilute gas limit of $r_0, r_n \ll r_1, r_5$. 
The size of this black hole is controlled by 
$gQ_1$ and $gQ_5$. 
Its thermodynamic quantities are given by
\begin{eqnarray}
M&=& { \pi \over 4 G_N^{5D}} \left [ 
r_1^2 + r_5^2 + {1 \over 2} r_0^2 \cosh 2\sigma \right ],
\nonumber \\ 
S &=& {{\cal A}_H^{5D} \over 4 G_N^{5D}} = 
{ \pi^2 r_1 r_5 r_0 \cosh \sigma \over 2 G_N^{5D}},
\label{thermo-quantity} \\
T_H &=& \left ( {2 \pi \over r_0} r_1 r_5 \cosh \sigma \right )^{-1}
\nonumber
\end{eqnarray}
with the 5D Newton constant ($G_N^{5D}$).
The above energy and entropy are essentially those of a gas of 
massless 1D particles. In this case the temperatures for left and 
right moving string modes are defined by
\begin{equation}
T_L = {1 \over 2 \pi} \left ( { r_0 \over r_1 r_5 } \right )e^{\sigma}, ~~
T_R = {1 \over 2 \pi} \left ( { r_0 \over r_1 r_5 } \right )e^{-\sigma}.
\label{temperatureLR}
\end{equation}
In the decoupling limit(near horizon, $r \simeq r_0$)  
the metric (\ref{string-metric}) leads to
\begin{eqnarray}
ds_{10}^2 &=& {r^2 \over R^2} \left ( -dt^2 + dx_5^2 \right ) +
{r_0^2 \over R^2} ( \cosh \sigma dt - \sinh \sigma dx_5)^2 
 \nonumber \\
&&
~~~~~~~~~~~~~~
+{R^2 \over r^2}  \left ( 1 - {r_0^2 \over r^2 } \right )^{-1}  dr^2 + 
R^2 d\Omega_3^2 + {r_1 \over r_5} dx_i^2
\label{dilute-metric}
\end{eqnarray}
with $R^2 = r_1 r_5$. Using the coordinate transformation 
$\rho^2 = r^2 + r_0^2 \sinh^2 \sigma$, one finds the 
well-known metric\cite{Hyu97,Sfe97,Alw98}
\begin{equation}
ds_{10}^2 = ds_{BTZ}^2 + R^2 d\Omega_3^2 + {r_1^2 \over R^2} dx_i^2,
\label{BTZ-metric}
\end{equation}
where the BTZ black hole space-time is given by\cite{Ban92}
\begin{equation}
ds_{BTZ}^2 = - {{(\rho^2 - \rho_+^2)(\rho^2 -\rho_-^2)} \over 
   {\rho^2 R^2 }} dt^2 + \rho^2 ( d \varphi - {J \over 2 \rho^2} dt )^2 +
  { \rho^2 R^2 \over {(\rho^2 - \rho_+^2)(\rho^2 -\rho_-^2)}} d \rho^2.
\label{BTZ-metric3}
\end{equation}
Here $M = (\rho_+^2 + \rho_-^2)/R^2, J=2 \rho_+ \rho_- /R$ are the mass and 
angular momentum of the BTZ black hole, and $\rho_+=r_0 \cosh \sigma , 
\rho_-=r_0 \sinh \sigma, \varphi = x_5 /R$. In this case, the relevant 
thermodynamic quantities(Hawking temperature, area of horizon, 
angular velocity at horizon, left/right temperatures) are given by
\begin{eqnarray}
T_H^{BTZ} &=& {\rho_+^2-\rho_-^2 \over 2 \pi R^2 \rho_+} = T_H,
\nonumber \\
{\cal A}_H^{BTZ} &=& 2 \pi \rho_+,~~ \Omega_H = { J \over 2 \rho_+^2},
\\ \label{temperature}
{1 \over T_{L/R}^{BTZ}} &=& {1 \over T_H^{BTZ}} 
    \left ( 1 \pm {\rho_+ \over \rho_-} \right ) = {1 \over T_{L/R}}.
\nonumber
\end{eqnarray}

\section{Supergravity Calculation : Scattering in 5D black hole}
We consider the scattering of a free scalar 
in the whole region of 5D black hole.
This means that our calculation corresponds to the bulk 
supergravity calculation. 
In the Einstein frame, $ds_{10}^2$ can be reduced 
to the 5D black hole and additional 
Kaluza-Klein moduli($\chi, \nu$) as\cite{Dow97} 
\begin{eqnarray}
\left ( ds_{10}^2 \right )_E &=& 
e^{-\phi/2} ds_{10}^2 
=e^{2 \chi} \sum_{i=6}^9 dx_i^2 +
  e^{2 \nu} ( dx_5 + A_\mu dx^\mu)^2 +
  e^{-2(4 \chi + \nu)/3} ds_{5D}^2
\label{kaluza-metric}
\end{eqnarray}
with $\mu = 0,1,2,3,4$.
Here $e^{2 \chi} = f_1^{1/4} f_5^{-1/4},~
e^{2 \nu} = f_1^{-3/4} f_5^{-1/4} f_n$,  the Kaulza-Klein 
gauge potential 
$A_0 = -r_0^2 \sinh 2 \sigma / 2(r^2+r_n^2)$, and 
the 5D black hole space-time is given by 
\begin{equation}
ds_{5D}^2 = - f^{-2/3} h dt^2 + f^{1/3} [ h^{-1} dr^2 + r^2 d \Omega_3^2]
\label{5d-metric}
\end{equation}
with $f = f_1 f_5 f_n$ with $f_n = 1 + r_n^2 / r^2$.

First let us consider the wave equation in the ten-dimensional background 
(\ref{kaluza-metric}),
\begin{equation}
\Box_{10} \Phi = 0.
\label{wave-eq}
\end{equation}
Considering the symmetries in (\ref{kaluza-metric}), $\Phi$ can be seperated 
as
\begin{equation}
\Phi = e^{-i \omega t} e^{i K_5 x_5} e^{i K_i x^i} Y_l(\theta_1, 
\theta_2, \theta_3) \phi(r).
\label{Phi}
\end{equation}
For simplicity, we take $K_5=0,~r_1 \simeq r_5$. 
Further we set $A_0=0$, because it approaches to 
zero in the dilute gas limit. Then (\ref{wave-eq}) leads to a 
new 5D wave equation for a free scalar
\begin{equation}
\left [ {d^2 \over dr^2} + \left ( { h' \over h} + {3 \over r} \right ) 
 { d \over dr} \right ] \phi + {\omega^2 f \over h^2} \phi 
- {l(l+2) \over h r^2} \phi - {K^2 f_5^2 \over h } \phi= 0.
\label{r-eq}
\end{equation}
For $l=K_i=0$ case, Eq.(\ref{r-eq}) corresponds to the 
wave equation of Ref.\cite{Dha96} and we recover that of 
Refs.\cite{Mat97,Mal97-1} for $K_i=0$. 
We confine ourselves to low energies satisfying $r_1 \omega, r_5 \omega < 1$. 
As usual we cannot find a solution to (\ref{r-eq}) analytically and thus 
we devide the space into a far region ($r \gg r_1, r_5$) and a near region 
($r \ll r_1, r_5$). We can match the solution in the overlapping 
region ($\omega r_0 \ll \omega r \ll 1$) because 
$\omega r_0 \ll \omega r_1, \omega r_5 < 1$ in the dilute gas limit
\cite{Dow97,Mat97,Mal97-1,Cal97}. 

In the far region we write $\phi = \tilde \phi/r$, and then the equation for 
$\phi$ leads to
\begin{equation}
{{d^2 \tilde \phi} \over d u^2} + 
{1 \over u} {d \tilde \phi \over d u} +
\left [ 1 - {\nu^2 \over u^2} \right ] \tilde \phi =0 
\label{far-eq}
\end{equation}
with $u=\omega' r$ and 
$\nu^2 = (l+1)^2 -(r_1^2 +r_n^2)K^2 -(\rho_1^2 +\rho_5^2 +\rho_n^2)$. 
Here $\omega' = \sqrt{\omega^2 -K^2}, \rho_i=\omega' r_i$ and 
$K_i r_i < \rho_i < \omega r_i$. 
The solutions are given by the Bessel function when $\nu$ is 
not an integer,
\begin{equation}
\phi_f = { 1\over u} \left [ \alpha J_\nu(u) +
   \beta J_{-\nu} (u) \right ],
\label{far-sol}
\end{equation}
where $\alpha, \beta$ are unknown constants.
From the large-$u$ behavior, one finds the incoming flux
\begin{eqnarray}
{\cal F}_{\rm in}(\infty)&=& { 2 \pi \over i} \left [ 
{\phi_f^{\rm in}}^* r^3 \partial_r \phi_f^{\rm in} - 
\phi_f^{\rm in} r^3 \partial_r {\phi_f^{\rm in}}^*
\right ] \bigg \vert_{r=\infty}
 \nonumber \\
&=& - { 2 \over \omega'^2} \left \vert 
\alpha e^{i(\nu + 1/2) \pi/2} +
\beta e^{i(-\nu + 1/2) \pi/2} 
\right \vert^2,
\label{far-flux}
\end{eqnarray}
where $\phi_f^{\rm in} $ is given by
\begin{equation}
\phi_f^{\rm in} = \sqrt{1 \over 2 \pi}
{ e ^{-i u } \over u^{3/2}} \left \{ 
\alpha e^{i(\nu + 1/2) \pi/2} +
\beta e^{i(-\nu + 1/2) \pi/2} 
\right \}.
\label{far-in-wave}
\end{equation}
The small-$u$ behavior of $\phi_f$ is 
\begin{equation}
\phi_{{\rm f} \to {\rm inter}} = { 1 \over u} \left [ 
\alpha \left ( { u \over 2} \right )^\nu { 1 \over \Gamma(\nu +1) } +
\beta \left ( { u \over 2} \right )^{-\nu} { 1 \over \Gamma(-\nu +1) } 
\right ].
\label{far-small-sol}
\end{equation}
Here we note that $\phi_{{\rm f} \to {\rm inter}}$ has not any pole 
because $\nu$ is not integer.

Now we wish to find the solution in the near region. Using 
$h= 1 - r_0^2 / r^2$, the wave equation (\ref{r-eq}) can be 
rewritten as
\begin{equation}
h(1-h) {d^2 \phi \over dh^2} + (1-h) {d \phi \over dh } +
\left \{ -C + {{C+D+E} \over h} + {E \over 1-h} \right \} \phi =0,
\label{near-eq}
\end{equation}
where
\begin{eqnarray}
C&=& \left ( { \omega r_1 r_5 r_n \over 2 r_0^2} \right )^2,
 \nonumber \\
D&=& {l(l+2) \over 4} + 
{ \omega^2 \over 4 r_0^2} \left [ r_1^2 r_5^2 + r_n^2(r_1^2 + r_5^2) 
\right ] + {K^2 r_5^2 \over 4} ,
 \nonumber \\
E&=& - {l(l+2) \over 4} + {\omega^2 (r_1^2+r_5^2 +r_n^2) \over 4} 
   - {K^2 (r_5^2-r_0^2) \over 4} .
\nonumber
\end{eqnarray}
Hereafter we set $K^2r_0^2, K^2 r_n^2 \simeq 0$. 
With an unknown constant $A$, we find the ingoing mode at the horizon
\begin{equation}
\phi^{\rm in}_n = A h^{-i q} (1-h)^{(1-\nu)/2} F(a,b,c;h),
\label{near-sol}
\end{equation}
where
\begin{eqnarray}
a &=& {1 -\nu \over 2} - i q + i \sqrt{C}, 
 \nonumber \\
b &=& {1 -\nu \over 2} - i q - i \sqrt{C}, 
 \nonumber \\
c &=& 1 - 2 i q,
 \nonumber \\
q &=& { \omega \over 4 \pi T_H} \sqrt{ ( 1 + {r_0^2 \over r_1^2} +
  {r_0^2 \over r_5^2} ) + 4 \pi^2 r_n^2 T_H^2}.
\nonumber
\end{eqnarray}
The large-$r$ behavior ($h\to 1$) follows from the ($h \to 1-h$) transformation 
law for hypergeometric functions as\cite{Abr66}
\begin{eqnarray}
\phi_{n \to f} &=& 
{{A \Gamma(1-2 i q) \Gamma(\nu)} \over 
   {\Gamma({{1+\nu}\over 2} -i q - i \sqrt{C})
   \Gamma({{1+\nu}\over 2} -i q + i \sqrt{C})}} u^{\nu -1}
 \nonumber \\
&& + ~~
{{A \Gamma(1-2 i q) \Gamma(-\nu)} \over 
   {\Gamma({{1-\nu}\over 2} -i q - i \sqrt{C})
   \Gamma({{1-\nu}\over 2} -i q + i \sqrt{C})}} u^{-\nu -1}.
\label{near-far-sol}
\end{eqnarray}
Matching (\ref{far-small-sol}) with (\ref{near-far-sol}) leads to 
\begin{eqnarray}
\alpha &=& A u_0 
{{ 2^\nu \Gamma(\nu+1) \Gamma(\nu)\Gamma(1 -2 i q) u_0^{-\nu}} \over 
   {\Gamma({{1+\nu}\over 2} -i q - i \sqrt{C})
   \Gamma({{1+\nu}\over 2} -i q + i \sqrt{C})}}, 
\\ \label{alpha-value}
\beta &=& A u_0 
{{ \Gamma(-\nu+1)\Gamma(-\nu)\Gamma(1 -2 i q) u_0^{\nu}} \over 
   {2^\nu \Gamma({{1-\nu}\over 2} -i q - i \sqrt{C})
   \Gamma({{1-\nu}\over 2} -i q + i \sqrt{C})}}. 
 \label{beta-value}
\end{eqnarray}
Since $u_0(=\omega' r_0) \ll 1$, one finds $ \beta \ll \alpha$.

Further the flux across the horizon is calculated as 
\begin{equation}
{\cal F}_{\rm in}(0) = - 8 \pi r_0^2 q | A |^2.
\label{near-flux}
\end{equation}
Then the absoprtion coefficient is given by
\begin{equation}
{\cal A} = {{\cal F}_{\rm in}(0) \over {\cal F}_{\rm in}(\infty)} \simeq
4 \pi u_0^2 q \left \vert {A \over \alpha} \right \vert ^2 .
\label{abs-coeff}
\end{equation}
The new absorption cross section takes the from
\begin{eqnarray}
\sigma_{\rm abs}^{5D} = (l+1)^2 { 4 \pi \over \omega^3}{\cal A} &=& 
\left({\omega' \over \omega} \right )^2 
{\cal A}_H^{5D} (\omega r_0)^{2 (\nu -1)} (l+1)^2
\left ( 1 + {r_0^2 \over r_1^2} +
  {r_0^2 \over r_5^2} + 4 \pi^2 r_n^2 T_H^2 \right ) ^{1/2}
 \nonumber \\
&& \times
\left \vert { 2^{-\nu +1} \over \Gamma(\nu) \Gamma(\nu+1) } \right \vert ^2 
\left \vert 
   {{\Gamma({{1+\nu}\over 2} -i q - i \sqrt{C})
   \Gamma({{1+\nu}\over 2} -i q + i \sqrt{C})} \over 
   \Gamma(1 -2 i q)
} 
\right \vert ^2.
\label{abs-cross}
\end{eqnarray}
The decay rate for $K\simeq 0, q \simeq \omega / 4 \pi T_H, \nu \simeq l+1$, is 
given by
\begin{eqnarray}
\Gamma_{5D} = {\sigma_{\rm abs}^{5D} \over { e^{\omega/T_H} -1}}
&=&
{{\pi r_0^{2 l+2} \omega^{2l-1}(l+1)^2 e^{-\omega/2T_H}} \over 
{\left \vert \Gamma(l+1) \Gamma(l+2)  \right \vert ^2}} 
\left \vert 
   \Gamma(1+{l\over 2} + i {\omega \over 2 \pi T_L})
   \Gamma(1+{l\over 2} + i {\omega \over 2 \pi T_R})
\right \vert ^2.
\label{decay-rate}
\end{eqnarray}
We now compare this decay rate with the CFT prediction. In the effective 
string picture(D-brane picture) such decays are described by 
a coupling of the space-time scalar field to an operator with 
dimension 1 in the conformal field theory, both in the left- and 
right-moving sector. A calculation showed that the 
decay rate is given by\cite{Mat97}
\begin{equation}
\tilde \Gamma_{5D} = {\Gamma_{5D} \over (l+1)^2}.
\label{decay-rate1}
\end{equation}
Hence the CFT prediction ($\tilde \Gamma_{5D}$) and the semiclassical 
decay rate ($\Gamma_{5D}$) agree at least when $l=0$.
Upto the first-order approximation(
$\omega' \simeq \omega ( 1- {K^2 \over 2 \omega^2}), 
q \simeq \omega / 4 \pi T_H, 
\nu \simeq l+1 - (K^2 + 2 \omega^2 )  r_5^2 / 2(l+1)$), we obtain the 
absorption cross section for $M_5 \times S^1$
\begin{eqnarray}
\left(\sigma_{\rm abs}^{6D}\right )^1  &=& 
\left ( 1 - {K^2 \over 2 \omega^2} \right ) ^2 
{\cal A}_H^{6D} \left (\omega\left (1-{(K^2+2 \omega^2) \over 2 \omega^2}\right ) 
r_0\right )^{2l -  (K^2+2 \omega^2) r_5^2 /(l+1)} (l+1)^2
\nonumber \\
&&~~~~\times
\left \vert { 2^{-l} \over 
\Gamma(l+1 - {(K^2+2 \omega^2) r_5^2 \over 2(l+1)}) 
\Gamma(l+2 - {(K^2+ 2 \omega^2) r_5^2 \over 2(l+1)}) }
\right \vert ^2 
 \nonumber  \\
&&~~~~ \times 
\left \vert 
{{\Gamma(1 + { l\over 2} - 
  { (K^2+2 \omega^2) r_5^2 \over 4(l+1)} - i { \omega \over 4 \pi T_L})
\Gamma(1 + { l\over 2} - 
  { (K^2+2 \omega62) r_5^2 \over 4(l+1)} - i { \omega \over 4 \pi T_R})}
\over 
\Gamma(1 - i {\omega \over 2 \pi T_H})
} 
\right \vert ^2.
\label{abs-cross1}
\end{eqnarray}
Here we used the dilute gas approximation
($r_1 \simeq r_5 \simeq R \gg r_0$) and 
${\cal A}_H^{6D} = {\cal A}_H^{5D} \times 2 \pi R = 4 \pi^3 r_n R^3$.
In this case one finds the relation 
(${\cal A}_H^{6D} T_H = 2 \pi^2 r_0^2 R$).
This is actually the same result with the geometry such as 
$AdS_3 \times S^3 \times T^4$ near horizon, and flat 
space-time at spatial infinity.

\section{$AdS$-Calculation : Scattering in $AdS_3 \times S^3 \times T^4$}
Here we consider the geometry such as the $AdS_3 \times S^3 \times T^4$ 
near horizon, 
and the flat space-time at spatial infinity.
A ten-dimensional minimally coupled scalar in the background 
of (\ref{BTZ-metric}) satisfies the wave equation
\begin{equation}
\Box_{10} \Psi = 0.
\label{BTZ-wave-eq}
\end{equation}
$\Psi$ can be decomposed into
\begin{equation}
\Psi = e^{-i \omega t} e^{i m \varphi} e^{i K_i x^i} 
  Y_l(\theta_1, \theta_2, \theta_3) \psi(\rho).
\label{psi-wave}
\end{equation}
Then (\ref{BTZ-wave-eq}) leads to 
\begin{equation}
\nabla^2_{BTZ} \psi(\rho) + {\mu \over R^2} \psi(\rho) =0,
\label{rho-eq}
\end{equation}
with $\mu = -l(l+2)-K^2 r_5^2$.
Since it is hard to find a solution to (\ref{rho-eq}) directly, 
we use the matching procedure\cite{Lee98}. 
The space is devided into two regions : 
the near region ($\rho \sim \rho_+$) and 
far region ($\rho \to \infty$). In the far region (\ref{rho-eq}) becomes 
\begin{equation}
{d^2 \psi_\infty \over dx^2} + { 3 \over x} { d \psi_\infty \over dx} +
 { \mu \over x^2} \psi_\infty =0 
\label{BTZ-far-eq}
\end{equation}
with a dimensionless quantity $x=\rho/R$.
Here one finds the far-region solution
\begin{equation}
\psi_\infty(x) = {1 \over x} \left [ \tilde \alpha x^{\sqrt{1-\mu}} +
   \tilde \beta x^{-\sqrt{1-\mu}} \right ].
\label{BTZ-far-sol}
\end{equation}
The ingoing flux at infinity is calculated as 
\begin{equation}
{\cal F}_{\rm in}(\infty) = 
- 2 \pi \sqrt{1-\mu} | \tilde \alpha - i \tilde \beta |^2.
\label{BTZ-far-flux}
\end{equation}
In order to obtain the near-region behavior, we introduce the 
variable $ z = (\rho^2 - \rho_+^2)/(\rho^2 - \rho_-^2) =
(x^2 - x_+^2)/(x^2 - x_-^2)$. Then (\ref{rho-eq}) becomes 
\begin{equation}
z(1-z) { d^2 \psi \over dz^2} + (1-z) { d \psi \over dz } 
+ \left ( { A_1 \over z} + { \mu/4 \over {1-z}} + B_1 \right ) \psi =0,
\label{BTZ-near-eq}
\end{equation}
where 
\begin{equation}
A_1 = \left ( { \omega - m \Omega_H \over 4 \pi T_H^{BTZ} } \right  ) ^2 ,
B_1 = - { \rho_-^2 \over \rho_+^2} \left ( 
{ \omega - m \Omega_H \rho_+^2/\rho_-^2 \over 4 \pi T_H^{BTZ}} \right ) ^2.
\nonumber 
\end{equation}
The ingoing solution for (\ref{BTZ-near-eq}) near horizon is given by 
the hypergeometric function
\begin{equation}
\psi(z) = C_1 z^{-i \sqrt{A_1}} (1 -z)^{(1 - \sqrt{1-\mu})/2} F(a,b,c;z),
\label{BTZ-near-sol}
\end{equation}
where 
\begin{eqnarray}
a&=& \sqrt{B_1} - i \sqrt{A_1} + (1 -\sqrt{1-\mu})/2,
 \nonumber  \\
b&=& -\sqrt{B_1} - i \sqrt{A_1} + (1 -\sqrt{1-\mu})/2,
 \nonumber  \\
c&=& 1 - 2 i \sqrt{A_1}.
\nonumber 
\end{eqnarray}
The corresponding flux is given by
\begin{equation}
{\cal F}_{\rm in}(0) = - 8 \pi \sqrt{A_1} ( x_+^2 - x_-^2) | C_1 |^2 .
\label{BTZ-near-flux}
\end{equation}
The absorption coefficient is calculated as
\begin{equation}
{\cal A} = {{\cal F}_{\rm in}(0) \over {\cal F}_{\rm in}(\infty) } =
{ 4 \sqrt{A_1} ( x_+^2- x_-^2) \over \sqrt{1-s}} 
{ |C_1|^2 \over | \tilde \alpha - i \tilde \beta |^2 } .
\label{BTZ-abs-coeff}
\end{equation}
To obtain $\tilde \alpha$ and $\tilde \beta$, we use the matching procedure.
First, we have to know the $z \to 1$ behavior 
of (\ref{BTZ-near-sol}). 
This can be obtained by the transformation rule ($ z \to 1-z$) 
for hypergeometric function\cite{Abr66}.  It takes the form 
\begin{equation}
\psi_{n \to f}(z) = {1 \over x} \left [ 
C_1 E_1 (x_+^2 - x_-^2)^{(1-\sqrt{1-\mu})/2} x^{\sqrt{1-\mu}} +
C_1 E_2 (x_+^2 - x_-^2)^{(1+\sqrt{1-\mu})/2} x ^{-\sqrt{1-\mu}} 
\right ],
\label{BTZ-near-far-sol}
\end{equation}
where
\begin{eqnarray}
E_1 &=& 
{\Gamma(1 -2 i \sqrt{A_1}) \Gamma(\sqrt{1-\mu}) \over 
 \Gamma({1 + \sqrt{1-\mu} \over 2} - \sqrt{B_1} -i \sqrt{A_1})
 \Gamma({1 + \sqrt{1-\mu} \over 2} + \sqrt{B_1} -i \sqrt{A_1})},
 \nonumber  \\
E_2 &=& 
{\Gamma(1 -2 i \sqrt{A_1}) \Gamma(-\sqrt{1-\mu}) \over 
 \Gamma({1 - \sqrt{1-\mu} \over 2} - \sqrt{B_1} -i \sqrt{A_1})
 \Gamma({1 - \sqrt{1-\mu} \over 2} + \sqrt{B_1} -i \sqrt{A_1})},
\nonumber
\end{eqnarray}
In matching (\ref{BTZ-far-sol}) with (\ref{BTZ-near-far-sol}), we find 
\begin{equation}
\tilde \alpha = C_1 E_1 \left ( { r_0 \over R} \right ) ^{1-\sqrt{1-\mu}},
~~~~\tilde \beta = C_1 E_2 \left ( { r_0 \over R} \right ) ^{1+\sqrt{1-\mu}}.
\label{BTZ-match}
\end{equation}
Considering $R(=\sqrt{r_1 r_5}) \gg r_0$, one finds 
$\tilde \beta \ll \tilde \alpha$.
Then the absorption coefficient is approximately given by 
\begin{equation}
{\cal A} \simeq { 4 \sqrt{A_1} (x_+^2 -x_-^2) \over \sqrt{1-\mu} }
\left ( { r_0 \over R} \right )^{2(\sqrt{1-\mu}-1)} 
{ 1 \over |E_1|^2}.
\label{BTZ-abs-coeff1}
\end{equation}
The absorption cross section for $AdS_3\times S^3$ with $m=0$ leads to
\begin{eqnarray}
\tilde \sigma_{\rm abs}^{6D} = c { {\cal A} \over \omega} &=& 
c { {\cal A}_H^{BTZ}\over \pi} 
\left ( {r_0 \over R} \right ) ^{2(\sqrt{1-\mu} -1)}
{ 1\over \Gamma(1+\sqrt{1-\mu})\Gamma(\sqrt{1-\mu}) }
 \nonumber  \\
&&\times
\left \vert 
{{\Gamma({1 + \sqrt{1-\mu} \over 2}  
   - i { \omega \over 4 \pi T_L^{BTZ}})
\Gamma({1 + \sqrt{1-\mu} \over 2}  
   - i { \omega \over 4 \pi T_R^{BTZ}})}
\over 
\Gamma(1 - i {\omega \over 2 \pi T_H^{BTZ}})
} 
\right \vert ^2.
\label{BTZ-abs-cross}
\end{eqnarray}
Here $c$ can be determined by requiring that $\lim_{\omega \to 0, s \to 0 } 
\tilde \sigma_{\rm abs}^{6D} = \tilde {\cal A}_H^{6D} = 
{\cal A}_H^{\rm BTZ} \times 2 \pi^2 R^3 $.
In this case the non-compact space-time is three dimensions, so that 
the conversion factor from ${\cal A}$ to $\sigma$ is $1/\omega$. 
Also one finds the relation : 
$\tilde {\cal A}_H^{6D} = {\cal A}_H^{6D} $.
Upto the first-order approximation 
($\sqrt{1-\mu} \simeq l+1 + {r_5^2 K^2 \over 2(l+1)}$),
 (\ref{BTZ-abs-cross}) leads to 
\begin{eqnarray}
\left ( \tilde \sigma_{\rm abs}^{6D}\right )^1 &=& 
\tilde {\cal A}_H^{6D} \left ( {r_0 \over R} \right )^{2l + {r_5^2 K^2 \over {l+1}}}
{ 1\over \Gamma(l+1 + {r_5^2 K^2 \over 2(l+1)})
         \Gamma(l+2 + {r_5^2 K^2 \over 2(l+1)}) }
 \nonumber \\
&&\times
\left \vert 
{{\Gamma(1 + {l \over 2}  
   + {r_5^2 K^2 \over 4(l+1)}
   - i { \omega \over 4 \pi T_L^{BTZ}})
\Gamma(1 + { l \over 2}  
   + {r_5^2 K^2 \over 4(l+1)}
   - i { \omega \over 4 \pi T_R^{BTZ}})}
\over 
\Gamma(1 - i {\omega \over 2 \pi T_H^{BTZ}})
} 
\right \vert ^2.
\label{BTZ-abs-cross1}
\end{eqnarray}

\section{Discussion}
We investigate the dynamical behavior of
the 5D black holes with a minimally coupled scalar field.
Apart from counting the microstates
of black holes, the dynamical behavior is also an important issue. 
This is so because the greybody factor(absorption cross section) for
the black hole arises as a consequence
of scattering of a free scalar off the 
potential surrounding the horizon.
That is, this is an effect of spacetime curvature. 

First we obtain the greybody factor for 5D black hole by using 
the supergravity calculation. 
Second, we calculate the absorption cross section of a
minimally coupled scalar by replacing the 5D black hole by
$AdS_3 \times S^3 \times T^4$.  
Here we do not require any boundary condition at infinity.
Actually we use the matching of $AdS$ region 
to obtain $\tilde \sigma_{\rm abs}^{6D}$.
In this sense our case is 
asymptotically anti-de Sitter space but not asymptotically flat space.
On the other hand, in studying the $AdS_3$, one usually requires the 
boundary conditions at spatial infinity\cite{Wit98,Lee98,Lif94,Bre82}. 
This is so because 
$AdS_3$ is not globally hyperbolic. The spatial 
infinity of the BTZ black hole is timelike, so the information 
may enter or exit from the bounadry at infinity. Thus one imposes boundary 
condition at infinity to obtain the sensible results.

Birmingham {\it et al.} showed that the s-wave greybody
factor for the BTZ black hole($AdS_3$) has the
same form as the one for 5D black hole($M_5$) in the dilute gas approximation
\cite{Bir97}.
In the low-energy approximation,
we find the same absorption cross section
for both two cases($M_5 \times S^1 \times T^4, AdS_3 \times S^3 \times T^4$) as
\begin{equation}
\left ( \sigma_{\rm abs}^{6D} \right )_{l=0,K=0} =
\left ( \tilde \sigma_{\rm abs}^{6D}\right )_{l=0,K=0}  \simeq
{\cal A}_H^{6D} 
\left \vert 
{{\Gamma(1 - i { \omega \over 4 \pi T_L})
\Gamma(1 - i { \omega \over 4 \pi T_R})}
\over 
\Gamma(1 - i {\omega \over 2 \pi T_H})
} 
\right \vert ^2.
\label{absorption}
\end{equation} 
This means that the relevant information of 5D black hole are encoded 
in the $AdS$-theory\cite{Str97,Alw98,Emp98}. 
Actually the greybody factor is determined by the potential 
surrounding the 5D black hole. If we introduce a tortoise 
coordinate $r_* = r + r_0 \ln | (r-r_0) / (r+r_0)|/2$
\cite{Dha96}, Eq.(\ref{r-eq}) can be rewritten as
\begin{equation}
\left [ - {d^2 \over dr_*^2} + V_\omega(r_*) \right ] \phi =0,
\label{eq-tortoise}
\end{equation}
where the $l=0$ potential $V_\omega$ is given by
\begin{equation}
V_\omega(r_*) = - \omega^2 f + K^2 f_5 h + {3 h \over 4 r^2} 
  \left ( 1 + {3 r_0^2 \over r^2} \right  ).
\label{potential}
\end{equation}
In the decoupling limit, (\ref{potential}) reduces to
\begin{equation}
V_\omega^{d-l}(r_*) = - \omega^2  {R^4 \over r^4} 
      \left ( 1 + {r_n^2 \over r^2} \right ) 
   + K^2 {R^2 \over r^2}  h + {3 h \over 4 r^2} 
  \left ( 1 + {3 r_0^2 \over r^2} \right  ),
\label{dl-potential}
\end{equation}
which corresponds actually to the potential for the 
$AdS_3 \times S^3 \times T^4$ near horizon. 
In the low-energy limit($\omega \to 0, K\to 0$), two potentials 
($V_\omega$ and $V_\omega^{d-l}$) take the same form. 
Hence it is clear that the absorption cross section of 
5D black hole is equal to that of the 
$AdS_3 \times S^3 \times T^4$ with asymptotically flat space.

In conclusion, we calculate the absorption cross section for a 
minimally coupled scalar in the $M_5 \times S^1 \times T^4$ and 
$AdS_3 \times S^3 \times T^4$ supergravities.  We note that 
the holography is related to theories of different 
domensionality\cite{Mal97,Gub98,Wit98}.  Here our working dimension 
is the same for two theories.  In the low-energy limit of $\omega \to 0$, two 
cross sections are same.  This shows a signal for holographic duality 
in the supergravity side.

Finally we comment on the greybody factors.  The calculations of 
greybody factor for a free scalar in $M_5$(5D black hole background) 
have already appeared in the literature\cite{Dha96,Mat97,Mal97-1}.  
In our calculation, some discrepancy appears because of the extra 
dimensions($T^4$).  For the $AdS_3$-calculation, Birmingham {\it et al.} 
first calculated the cross section for a free scalar\cite{Bir97}.  
Although this is correct, the calculation procedure is incomplete.  This is so 
because they did not consider the boundary condition at infinity seriously.  
One cannot obtain the cross section for a free scalar in the $AdS_3$ by 
requiring the Dirichlet or Neumann boundary condition at spatial infinity.  
Our calculation for $AdS_3 \times S^3\times T^4$ is done with asymptotically 
$AdS_3$.  However we do not need to consider the boundary condition.  
Actually this is based on the non-normalizable 
mode ($\tilde \alpha$ - dependent term in (\ref{BTZ-far-sol})) 
without the boundary condition\cite{Wit98,Bal98}.
Even in the asymptotically $AdS_3$, it seems that one can calculate 
the absorption cross section in the $AdS_3$ by following our procedure.

\section*{Acknowledgement}
This work was supported in part by the Basic Science Research Institute 
Program, Minstry of Education, Project NOs. BSRI-97-2413 and 
BSRI-97-2441.

\end{document}